\documentclass{aastex}          
\usepackage{spr-astr-addons}    
\begin{document}

\title{The Cosmic Equation of State}

\shorttitle{The Cosmic Equation of State}
\shortauthors{Melia}

\author{F. Melia\altaffilmark{1}}
\affil{Department of Physics, the Applied Math Program, and Department of Astronomy,
The University of Arizona, Tucson, AZ 85721 \\
E-mail: fmelia@email.arizona.edu}

\altaffiltext{1}{John Woodruff Simpson Fellow.} 

\begin{abstract}
The cosmic spacetime is often described in terms of the Friedmann-Robertson-Walker
(FRW) metric, though the adoption of this elegant and convenient solution to Einstein's
equations does not tell us much about the equation of state, $p=w\rho$, in terms of the
total energy density $\rho$ and pressure $p$ of the cosmic fluid. $\Lambda$CDM and
the $R_{\rm h}=ct$ Universe are both FRW cosmologies that partition $\rho$ into
(at least) three components, matter $\rho_{\rm m}$, radiation $\rho_{\rm r}$, and a
poorly understood dark energy $\rho_{\rm de}$, though the latter goes one step
further by also invoking the constraint $w=-1/3$. This condition is apparently required
by the simultaneous application of the Cosmological principle and Weyl's postulate.
Model selection tools in one-on-one comparisons between these two cosmologies favor
$R_{\rm h}=ct$, indicating that its likelihood of being correct is $\sim 90\%$
versus only $\sim 10\%$ for $\Lambda$CDM. Nonetheless, the predictions of
$\Lambda$CDM often come quite close to those of $R_{\rm h}=ct$, suggesting
that its parameters are optimized to mimic the $w=-1/3$ equation-of-state.
In this paper, we explore this hypothesis quantitatively and demonstrate 
that the equation of state in $R_{\rm h}=ct$ helps us to understand why the
optimized fraction $\Omega_{\rm m}\equiv \rho_m/\rho$ in $\Lambda$CDM
must be $\sim 0.27$, an otherwise seemingly random variable. We show that
when one forces $\Lambda$CDM to satisfy the equation of state
$w=(\rho_{\rm r}/3-\rho_{\rm de})/\rho$, the value of the Hubble radius today,
$c/H_0$, can equal its measured value $ct_0$ only with $\Omega_{\rm m}\sim0.27$
when the equation-of-state for dark energy is $w_{\rm de}=-1$. (We also show, however,
that the inferred values of $\Omega_{\rm m}$ and $w_{\rm de}$ change in a correlated fashion
if dark energy is not a cosmological constant, so that $w_{\rm de}\not= -1$.)
This peculiar value of $\Omega_{\rm m}$ therefore appears to be a direct
consequence of trying to fit the data with the equation of state
$w=(\rho_{\rm r}/3-\rho_{\rm de})/\rho$ in a Universe whose principal
constraint is instead $R_{\rm h}=ct$ or, equivalently, $w=-1/3$.
\end{abstract}


\section{Introduction}
The Cosmological principle and Weyl's postulate appear to be essential
ingredients in any physically realistic cosmological theory. Together,
they posit that the Universe is homogeneous and isotropic (at least on large,
i.e., $>100$ Mpc, spatial scales), and that this high degree of symmetry is
maintained from one time slice to the next. The appropriate spacetime to use
is conveniently and elegantly written in terms of the Friedmann-Robertson-Walker
(FRW) metric though this, in and of itself, does not tell us much about the
cosmic equation of state, relating the total energy density $\rho$ to
its total pressure $p$.

In principle, if we knew these quantities precisely, we could then solve the
dynamical equations governing the Universal expansion and understand its
large-scale structure and how it evolved to its current state.  One could then
also unambiguously interpret many of the observations, including the redshift-dependent
luminosity distance to Type Ia SNe and the spectrum of fluctuations in the cosmic
microwave background (CMB). Unfortunately, we must rely on measurements and
intuition to pick $\rho$ and $p$. The best we can do today is to assume that
$\rho$ must contain matter $\rho_{\rm m}$ and radiation
$\rho_{\rm r}$, which we see directly, and an as yet poorly understand
`dark' energy $\rho_{\rm de}$, whose presence is required by a broad range
of data including, and especially, the aforementioned Type Ia SNe
\citep{Riess1998,Perlmutter1999}. But instead of refining the cosmic equation
of state, $p=w\rho$, the ever-improving measurements of the redshift-distance
and redshift-age relations seem to be creating more tension between theory
and observations, rather than providing us with a better indication of the
dark-energy component, $p_{\rm de}=w_{\rm de}\rho_{\rm de}$. For the
other two constituents, one simply uses the prescription $p_{\rm r}=\rho_{\rm r}/3$
and $p_{\rm m}\approx 0$, consistent with a fully relativistic fluid
(radiation) on the one hand, and a non-relativistic fluid (matter) on the other.

One of the most basic FRW models, $\Lambda$CDM, assumes that dark energy is a
cosmological constant $\Lambda$ with $w_{\rm de}\equiv w_\Lambda=-1$, and
therefore $w=(\rho_{\rm r}/3-\rho_\Lambda)/\rho$. This model does quite
well explaining many of the observations, but growing empirical evidence
suggests that it is inadequate to explain all of the nuances
seen in cosmic evolution and the growth of structure. For example, $\Lambda$CDM
cannot account for the general uniformity of the CMB across the sky
without invoking an early period of inflated expansion \citep{Guth1981,Linde1982},
yet the latest observations with {\it Planck} \citep{Ade2013} suggest that
the inflationary model may be in trouble at a fundamental level
\citep{Ijjas2013,Ijjas2014,Guth2013}. Insofar as the CMB fluctuations
measured with both WMAP \citep{Bennett2003} and {\it Planck} are concerned,
there appears to be unresolvable tension between the predicted and measured
angular correlation function \citep{Copi2009,Copi2013,Melia2014a,Bennett2013}.
And there is also an emerging conflict between the
observed matter distribution function, which is apparently scale-free, and
that expected in $\Lambda$CDM, which has a different form on different spatial
scales. The fine tuning required to resolve this difference led Watson et al.
\citep{Watson2011} to characterize the matter distribution function as a `cosmic coincidence.'
Such difficulties are compounded by $\Lambda$CDM's predicted redshift-age
relation, which does not appear to be consistent with the growth of quasars
at high redshift \citep{Melia2013a}, nor the very early appearance of galaxies at
$z\gtrsim 10$ \citep{Melia2014b}.

It is therefore important to refine the basic $\Lambda$CDM
model, or perhaps to eventually replace it if necessary, to improve the
comparison between theory and observations. Over the past several years,
we have been developing another FRW cosmology, known as the $R_{\rm h}=ct$
Universe, that has much in common with $\Lambda$CDM, but includes an additional
ingredient motivated by several theoretical and observational arguments
\citep{Melia2007,MeliaAbdelqader2009,MeliaShevchuk2012}.
Like $\Lambda$CDM, it also adopts the equation of state $p=w\rho$, with $p=p_{\rm m}+
p_{\rm r}+p_{\rm de}$ and $\rho=\rho_{\rm m}+\rho_{\rm r}+\rho_{\rm de}$, but
is subject to the additional constraint that $w=(\rho_{\rm r}/3+
w_{\rm de}\rho_{\rm de})/\rho=-1/3$ at all times. One might come away with the
impression that these two prescriptions for the equation of state cannot be
consistent. But in fact if we ignore the constraint $w=-1/3$ and instead
proceed to optimize the parameters in $\Lambda$CDM by fitting the data, the
resultant value of $w$ averaged over a Hubble time  is actually $-1/3$ within the measurement errors
\citep{Melia2007,MeliaShevchuk2012}. In other words, though $w=(\rho_{\rm r}/3-\rho_\Lambda)/\rho$
in $\Lambda$CDM cannot be equal to $-1/3$ from one moment to the next, its value averaged
over the age of the Universe is equal to what it would have been in $R_{\rm h}=ct$ anyway.

This result does not necessarily prove that $\Lambda$CDM is an incomplete version of
$R_{\rm h}=ct$, but it does seem to suggest that the inclusion of the additional
constraint $w=-1/3$ might render its predictions closer to the data. By now,
comparative analyses of $\Lambda$CDM and $R_{\rm h}=ct$ have been carried out
for a broad range of observations, from the CMB \citep{Melia2014a}, high-$z$ quasars
\citep{Melia2013a,Melia2014b} and the ages of high-$z$ objects \citep{Melia2014b,Yu2014} 
in the early Universe, to gamma-ray bursts \citep{Wei2013a} and
cosmic chronometers \citep{MeliaMaier2013} at intermediate redshifts and, most recently,
to the relatively nearby Type Ia SNe \citep{Wei2013b}. In every case,
model selection tools indicate that the likelihood of $R_{\rm h}=ct$ being correct is
typically $\sim 90\%$ compared with only $\sim 10\%$ for $\Lambda$CDM. And perhaps
the most important distinguishing feature between these two cosmologies is that, whereas
$\Lambda$CDM cannot survive without inflation, the $R_{\rm h}=ct$ Universe
does not need it in order to avoid the well-known horizon problem \citep{Melia2014c}.
Thus, an eventual abandonment of inflation should it fail to work
self-consistently would completely tip the scale in favor of $R_{\rm h}=ct$.

The purpose of this paper is to further develop the $R_{\rm h}=ct$
Universe by addressing a rather obvious question that comes to mind. Since $\Lambda$CDM
lacks the ingredient $w=-1/3$ that would turn it into $R_{\rm h}=ct$, why does it
in fact do quite well in accounting for many of the data? And are there any other
obvious observational consequences of the prescription $w=(\rho_{\rm r}/3-
\rho_\Lambda)/\rho$ for its equation of state? Here, we demonstrate
that the inclusion of the condition $w=-1/3$ in $\Lambda$CDM actually
helps to explain why the fraction $\Omega_{\rm m}\equiv \rho_m(t_0)/\rho(t_0)$
of its energy density in the form of (visible and dark) matter today must be
$\approx 0.27$ in order for it to adequately fit the data. In other words, we will
show that the inferred value of $\Omega_{\rm m}$ in
$\Lambda$CDM is not random at all, but is instead uniquely required when one
attempts to account for the observations using the equation of state $w=(\rho_{\rm r}/3-
\rho_\Lambda)/\rho$ in a Universe that is in reality evolving according to the
constraint $w=(\rho_{\rm r}/3+w_{\rm de}\rho_{\rm de})/\rho=-1/3$.
We will demonstrate this interesting and important connection between
$\Lambda$CDM and $R_{\rm h}=ct$  in \S\S~2 and 3, and discuss the
results in \S~4.

\section{The Cosmic Spacetime}
The basic $\Lambda$CDM model avoids having to deal with uncertainties
in the particle physics by relying on transitions, starting with 
an early radiation-dominated phase, followed by a Universe dominated by matter
after recombination, and then transitioning into a period dominated by
dark energy. But in order to make testable predictions, we have to
assume values for $\Omega_{\rm m}$, $\Omega_{\rm r}$ and $\Omega_\Lambda$,
and then integrate backwards to the big bang by solving the dynamics
equations using the equation of state $w=(\rho_{\rm r}/3-\rho_\Lambda)/\rho$.
If the Universe is truly subject to the constraint $w=-1/3$ at all times,
however, how does this affect the observable signatures and inferred parameters
of the standard model? This is the question we will now attempt to answer.

The Friedmann-Robertson-Walker metric may be written
\begin{eqnarray}
ds^2 &=& c^2 dt^2 - a^2(t)[dr^2 (1 - kr^2)^{-1} + \nonumber \\
&\null& \qquad\qquad\qquad\qquad r^2(d\theta^2 + \sin^2\theta d\phi^2)]\;,
\end{eqnarray}
in terms of the cosmic time $t$ in the comoving frame, and the corresponding 
radial ($r$) and angular ($\theta$ and $\phi$) coordinates. 
The expansion factor $a(t)$ is a function of $t$ only, whereas the spatial coordinates
$(r,\theta,\phi)$ in this frame remain ``fixed" for all particles in the cosmos. The constant
$k$ is $+1$ for a closed universe, $0$ for a flat, open universe, or $-1$ for an open universe.

The source of spacetime curvature in a Universe that satisfies the Cosmological Principle is
a perfect fluid \cite{Weinberg1972} which, together with the metric coefficients appearing
in Equation~(1), allows us to simplify Einstein's equations
and derive the key dynamical expressions governing the smoothed-out expansion
at large scales. These include, respectively, the Friedmann and energy-conservation equations,
\begin{equation}
H^2\equiv\left({\dot a\over a}\right)^2={8\pi G\over 3c^2}\rho-{kc^2\over a^2}\;,
\end{equation}
and
\begin{equation}
\dot\rho=-3H(\rho+p)\;,
\end{equation}
both written in terms of the total energy density $\rho$ and total pressure $p$.
$H$ is the time-dependent Hubble `constant' and an overdot denotes a derivative with
respect to time $t$.

In the $R_{\rm h}=ct$ Universe, the `active mass' is zero, meaning that 
$\rho+3p=0$ \citep{Melia2014d}. Therefore, from the definition of the gravitational
radius $R_{\rm h}=2GM/c^2$, in terms of the Misner-Sharp mass $M=(4\pi/3)R^3_{\rm h}
(\rho/c^2)$ \citep{Misner1964}, it is easy to show that ${\dot{R}}_{\rm h}=
(3/2)(1+w)c$, where $w=p/\rho$ \citep{MeliaShevchuk2012}, which yields
$R_{\rm h}=ct$ (the eponymous constraint of this model). And since the 
gravitational radius $R_{\rm h}$ is a proper distance in this spacetime, one 
must also have $H=1/t$ (see, e.g., Melia \& Shevchuk 2012) which, 
together with Equation~(2), then shows that $k=0$. 

Interestingly the CMB strongly constrains the total energy density to be 
near its critical value, $\rho_c\equiv 3c^2H_0^2/8\pi G$, where $H_0\equiv H(t_0)$ 
\citep{Bennett2003,Spergel2003}, so the observations appear to be consistent
with zero spatial curvature. Though this empirical result emerges from the
optimization of model parameters in $\Lambda$CDM, the fact that the standard
model is often a good approximation to $R_{\rm h}=ct$ lends some observational
support for this theoretical prediction of the $R_{\rm h}=ct$ cosmology. For these
reasons, it will be sensible for us to assume a perfectly flat universe, and we will here 
always assume that $k=0$. This also means that $\Omega\equiv \Omega_{\rm r}+
\Omega_{\rm m}+\Omega_{\rm de}=1$. Our analysis in this paper will be
based entirely on this premise. It is therefore straightforward to integrate
Equation~(2), yielding
\begin{equation}
ct_0=R_{\rm h}(t_0)\,\int_0^1 {u\,du\over \sqrt{\Omega_{\rm r}+\Omega_{\rm m}u+
\Omega_{\rm de}u^{1-3w_{\rm de}}}}\;.
\end{equation}
To obtain this expression, we have allowed for the possibility that dark energy is
not a cosmological constant (i.e., that $w_{\rm de}$ may be different from $-1$,
in which case we would refer to this model as wCDM, rather than $\Lambda$CDM),
and we have used the derived value of the gravitational horizon to write $R_{\rm h}=c/H$
\citep{Melia2007,MeliaShevchuk2012}. This expression also assumes that $a\rightarrow 0$ at $t=0$.

Equation~(4) must be satisfied by every flat FRW cosmology, though
the explicit dependence of the integrand on $\Omega_{\rm m}$, $\Omega_{\rm r}$,
and $\Omega_{\rm de}$ shown here applies specifically to $\Lambda$CDM
(or wCDM if $w_{\rm de}\not= -1$).
However, if in fact the Cosmological Principle and Weyl's postulate require the
equation of state $w=-1/3$, then $R_{\rm h}(t_0)=ct_0$ \citep{Melia2007,MeliaShevchuk2012},
so $\Lambda$CDM (or wCDM) would have no choice but to satisfy the condition
\begin{equation}
{\mathcal{I}}\equiv \int_0^1 {u\,du\over \sqrt{\Omega_{\rm r}+\Omega_{\rm m}u+
\Omega_{\rm de}u^{1-3w_{\rm de}}}}=1\;.
\end{equation}
Let us now see what the consequences of this constraint are for $\Lambda$CDM.
Figure~1 shows the calculated value of this integral ${\mathcal{I}}$ as a function
of $\Omega_{\rm m}$, for various dark-energy equations of state, $w_{\rm de}$.
The radiation energy density is evaluated on the basis of the CMB's current
temperature, $T=2.7$ K. Not surprisingly, ${\mathcal{I}}$
can have a broad range of values, but for any given $w_{\rm de}$, there
is only one unique determination of $\Omega_{\rm m}$ that satisfies the
condition ${\mathcal{I}}=1$. And for the special case of a cosmological
constant ($w_{\rm de}=-1$), that value is $0.27$.

\section{Observational Constraints}
Over the past decade, both $\Omega_{\rm m}$ and $w_{\rm de}$ have been
measured with relatively high precision, combining constraints from
a variety of observational data sets. For example, \cite{Melchiorri2003}
combined data from six CMB experiments \citep{Spergel2003}, from the power
spectrum  of large-scale structure in the 2dF 100k galaxy redshift survey
\citep{Tegmark2002}, from luminosity measurements of Type Ia SNe
\citep{Riess1998,Perlmutter1999}, and from the Hubble space telescope
measurements of the Hubble parameter $H_0$. More recent
analyses have refined the quantitative results from this extensive
survey, though not altering the basic conclusions. It is therefore
rather straightforward for us to compare our theoretical
predictions directly with the observations. As we shall see shortly,
the story emerging from this exercise is quite revealing.

None of the individual observations results in fits that are so precise
as to produce unique values for the parameters $(w_{\rm de},\Omega_{\rm m})$.
The reason for this is that, other than the Sachs-Wolfe effect \citep{Sachs1967},
which is responsible for the largest angular
fluctuations in the CMB, none of the other mechanisms producing
structure of one kind or another depends sensitively on the expansion
history of the Universe. As such, some degeneracy exists among the
possible choices of cosmological parameters pertaining to the CMB
\citep{Kosowsky2002}.

At lower redshifts, the cosmological measurements are heavily influenced
by the observation of Type Ia SNe. But here also, both the luminosities
and angular distances (the fundamental observables) depend on
$w_{\rm de}$ through multiple integrals, and are therefore not particularly
sensitive to variations in $w_{\rm de}$ with redshift \citep{Maor2001}.

\begin{figure}[h]
\vskip 0.2in
\includegraphics[width=\columnwidth]{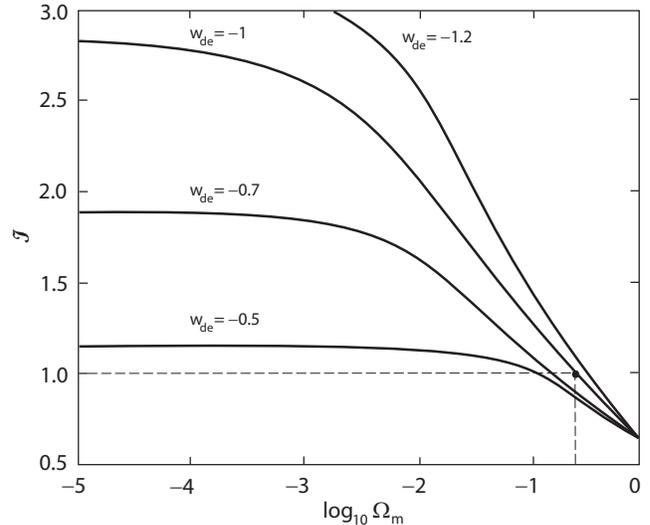}
\caption{The ratio $\mathcal{I}\equiv ct_0/R_{\rm h}(t_0)$, calculated as a function
of $\Omega_{\rm m}$, according to $\Lambda$CDM (or wCDM when $w_{\rm de}\not=-1$). 
The label $w_{\rm de}$ indicates the
corresponding equation of state for dark energy, i.e., $p_{\rm de}=w_{\rm de}\rho_{\rm de}$,
in terms of its pressure $p_{\rm de}$ and density $\rho_{\rm de}$. When
$R_{\rm h}(t_0)=ct_0$ and $w_{\rm de}=-1$, $\Omega_{\rm m}$ must have the
unique value $0.27$ (indicated by the black dot).}
\end{figure}

Nonetheless, all of the constraints derived from the
various data sets do produce a well-defined region in $w_{\rm de}-
\Omega_{\rm m}$ phase space where the most likely values of these
parameters are expected to be found. The confidence regions
shown in figure~2 are adapted from a corresponding figure in
\cite{Melchiorri2003}. These show the $68\%$, $95\%$, and
$99\%$ confidence regions corresponding to the Type Ia SNe
observations (adapted from Suzuki et al. 2012, 
shown as gray swaths), and the corresponding
regions inferred from the analysis of CMB, HST, and 2dF data
(indicated by the lighter-colored island regions to the
upper left of this diagram). Insofar as the values of $w_{\rm de}$
and $\Omega_{\rm m}$ are concerned, the supernova data are not
as constraining as the other sets, but there is clearly a satisfying
consistency among all of the observations.

Also shown in figure~2 is one of the more interesting results
of this paper, indicated here as a thick black curve to the left of
this diagram. This feature shows the loci of $(w_{\rm de},
\Omega_{\rm m})$ points permitted by the requirement that
the integral ${\mathcal{I}}$ be equal to 1 (see also
figure~1). That is, while the constraints shown in figure~2
are based on the interpretation of the data using $\Lambda$CDM,
this theoretical curve goes one step further, by illustrating what
values of $w_{\rm de}$ and $\Omega_{\rm m}$ are actually permitted
theoretically when we impose the additional constraint
$R_{\rm h}(t_0)=ct_0$ (or, equivalently, the equation of state
$w=-1/3$).  Notice, for example, where the latest measurement
of $\Omega_{\rm m}$ and $w_{\rm de}$ with Planck fall on this
diagram (the star in figure~2). Whereas $\Omega_{\rm m}=0.27$
is linked to a dark-energy equation of state $w_{\rm de}=-1$, the
Planck measurement of $\Omega_{\rm m}\approx 0.3$ is associated
with $w_{\rm de}=1.13^{+0.13}_{-0.10}$ \citep{Ade2013}.

On its own, $\Lambda$CDM has
no explanation for why the most preferred region of allowed
values is limited to $-1.38<w_{\rm de}<-0.82$ and $0.22<
\Omega_{\rm m}<0.35$, and why this oblong region is slanted
in such a way as to couple the higher values of $w_{\rm de}$ to
the smaller values of $\Omega_{\rm m}$. But in the context of $R_{\rm h}
=ct$, this is precisely the region permitted by the requirement
that ${\mathcal{I}}$ be equal to 1, as evidenced by the fact
that our theoretical curve passes directly through the middle
of the observationally permitted region and, even more
impressively, precisely tracks the {\it orientation} of this region.
The point of this is that while the data are not sufficiently
precise to tell us the exact value of $\Omega_{\rm m}$, the range
of allowed values of $w_{\rm de}$ trends with $\Omega_{\rm m}$ in such
a way as to always preserve the condition ${\mathcal{I}}=1$.

\begin{figure}[h]
\includegraphics[width=\columnwidth]{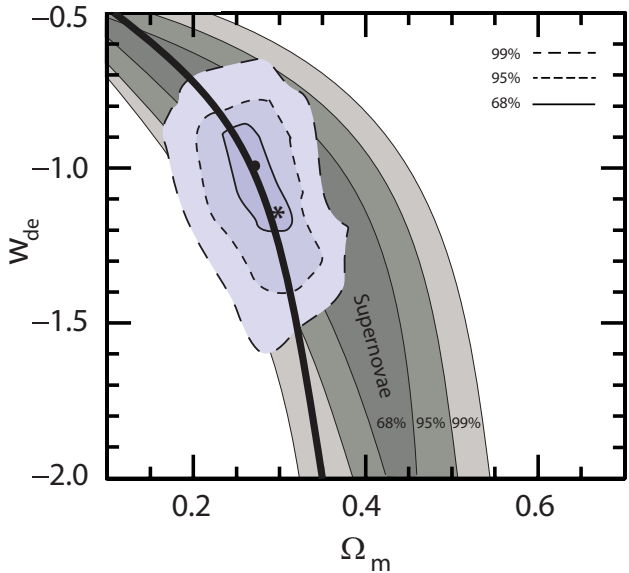}
\caption{The solid black curve indicates the value $w_{\rm de}$
must have in $\Lambda$CDM (or wCDM if $w_{\rm de}\not=-1$)
as a function of $\Omega_{\rm m}$, when the condition $R_{\rm h}(t_0)=ct_0$ 
is imposed. This curve is shown against the constraints (adapted from 
figure~4 in Melchiorri et al. 2003) on the dark-energy equation-of-state, 
assuming a flat universe. The Type Ia SN limits have been updated from
the more recent results in Suzuki et al. (2012). These limits and 
confidence levels include results from CMB anisotropies, measurements of the 
Hubble constant, and large-scale structure. The empirically derived, 
concordance values of $w_{\rm de}$ versus $\Omega_{\rm m}$ track those
imposed on $\Lambda$CDM by the $R_{\rm h}(t_0)=ct_0$ condition exceptionally well.
Note, for example, the location (black dot) of the WMAP measurements 
\citep{Bennett2013}, versus (star) the latest measurements by Planck \citep{Ade2013}, 
which resulted in the values $\Omega_{\rm m}\approx 0.3$ and $w_{\rm de}\approx 
-1.13$. The value $\Omega_{\rm m}=0.27$ is realized only when $w_{\rm de}=-1$.}
\end{figure}

These results clearly argue against any suggestion that
$\Omega_{\rm m}\sim 0.3$ (or, more specifically, $\Omega_{\rm m}=0.27$
when $w_{\rm de}=-1$) could be a coincidence in $\Lambda$CDM. First,
not only would it be highly improbable for $\Omega_{\rm m}$ to have
the value required to guarantee $R_{\rm h}(t_0)=ct_0$
which, by the way, could only happen once in the entire
history of the Universe, and it would have to be happening
right now, when we just happen to be looking. But in addition,
the region of $w_{\rm de}-\Omega_{\rm m}$ phase space permitted
by the data shows a clear trend exactly matching the
behavior one would expect if ${\mathcal{I}}$ must always
be equal to 1. In other words, even if $\Omega_{\rm m}\sim 0.3$
were somehow a coincidence, there is no reason why the
allowed region of $w_{\rm de}-\Omega_{\rm m}$ phase space should
be slanted from upper left to bottom right, instead of from
upper right to bottom left.

It is therefore difficult to argue against the conclusion that
$\Lambda$CDM is merely mimicking the expansion history
we would have obtained with $R_{\rm h}=ct$ all along, and
that the observed value of $\Omega_{\rm m}$ (which happens to
be $0.27$ if $w_{\rm de}=-1$) is required in order to make the
assumed density $\rho=\rho_r+\rho_m+\rho_{\rm de}$ comply
with the equation of state $p=-\rho/3$ found in the $R_{\rm}=
ct$ Universe.

\section{Discussion and Conclusion}
The results we have just presented do not exist in isolation, of course.
They add weight to the other one-on-one comparisons between
$R_{\rm h}=ct$ and $\Lambda$CDM that uniformly show the
superiority of the former over the latter in accounting for the data.
But the analysis we have carried out in this paper is important specifically
because it starts to probe the fundamental reasons why $\Lambda$CDM
can sometimes function as an approximation to $R_{\rm h}=ct$, and why it does
reasonably well accounting for some of the data, e.g., the Type Ia SNe.
For example, even though the empirically motivated choice of density
$\rho=\rho_r+\rho_m+\rho_{\rm de}$ is not entirely consistent with the
equation of state $p=-\rho/3$, it can nonetheless lead to an expansion
history that mimics $R_{\rm h}=ct$ over a Hubble time---but only so
long as $\Omega_{\rm m}\sim 0.27$.

Recently, we studied in detail how the Type Ia SNe ought to be interpreted
in the context of $\Lambda$CDM and $R_{\rm h}=ct$ \citep{Melia2012a,Wei2013b}.
The best-fit distance moduli calculated from these two
theories are so close to each other all the way out to $z\sim 6$, that it is
difficult to determine on the basis of a $\chi^2$ comparison alone which
of these two cosmologies is favored. This is due in part to the
strong dependence of the data reduction itself on the pre-assumed
cosmology, since at least 4 `nuisance' parameters defining the SN
luminosity must be optimized along with the free parameters of the
model. The inferred SN luminosities and their `measured' distance
moduli are therefore strongly compliant to the pre-assumed model,
greatly weakening this particular comparative test. Indeed, a
similar analysis of the most up-do-date Gamma-ray Burst Hubble
Diagram (HD) \citep{Wei2013a} reinforces this point by demonstrating
that when the data are re-calibrated correctly for each individual
cosmology, the $R_{\rm h}=ct$ Universe fits the observed HD
better than $\Lambda$CDM does.

A quick inspection of figure~3 allows us to better appreciate why
$\Lambda$CDM fits the Type Ia SNe and Gamma-ray
Burst data as well as it does. This figure shows
the ratio of luminosity distances $d_L^{\Lambda{\rm CDM}}/d_L^{R_{\rm h}=ct}$
as a function of redshift for different values of $\Omega_{\rm m}$, in
a Universe with $w_{\rm de}=-1$. What emerges from this diagram is
that the value of $\Omega_{\rm m}$ that comes closest to satisfying the
condition $R_{\rm h}(t_0)=ct_0$ in Equation~(4), also corresponds
to the $\Lambda$CDM universe in which the luminosity distance
$d_L^{\Lambda{\rm CDM}}$ most closely tracks its counterpart in
$R_{\rm h}=ct$. One should not be surprised therefore, to see
that the best fit $\Lambda$CDM cosmology fits the Type Ia SNe
and Gamma-ray Burst data as well as it does.

\begin{figure}[h]
\includegraphics[width=\columnwidth]{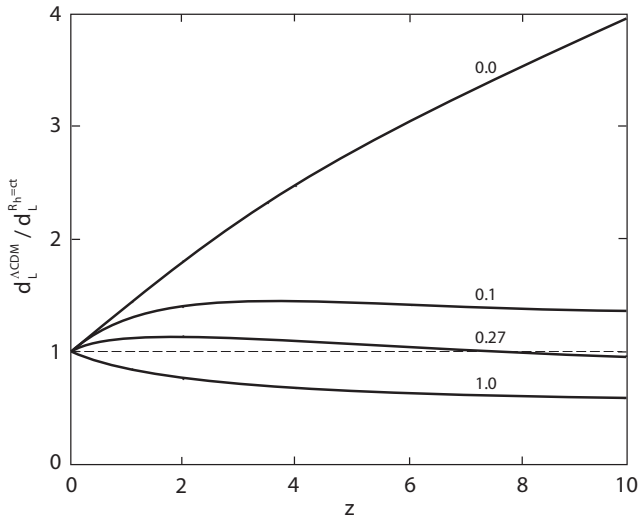}
\caption{Ratio of luminosity distance in $\Lambda$CDM over that
in $R_{\rm h}=ct$, as a function of redshift, for various values
of $\Omega_{\rm m}$, assuming $w_{\rm de}=-1$. The $\Lambda$CDM cosmology
with $\Omega_{\rm m}=0.27$, which comes closest to satisfying the condition
$R_{\rm h}(t_0)=ct_0$, also best approximates the
condition $d_L^{\Lambda{\rm CDM}}=d_L^{R_{\rm h}=ct}$ over a
large range in $z$.}
\end{figure}

Unfortunately, $\Lambda$CDM does not do as well accounting
for the high-$z$ universe, having difficulty explaining why the CMB fluctuations
show no correlation at angles greater than $\sim 60^\circ$, and
failing to explain how $\sim 10^9\, M_\odot$ supermassive black holes
could have formed so quickly after the big bang. The issue is that
even though the differences between $\Lambda$CDM and $R_{\rm h}
=ct$ may be smoothed out over a Hubble time through the careful choice
of $\Omega_{\rm m}\sim 0.27$, the expansion history of the Universe at
$z> 6$ is so different between these two cosmologies that it is simply
not possible to mimic the equation of state $p=-\rho/3$ with
$\rho=\rho_r+\rho_m+\rho_{\rm de}$, given that $R_{\rm h}=ct$
predicts a constant expansion, while $\Lambda$CDM predicts
a very rapid deceleration. Additional inconsistencies between 
the predicted age-redshift relationship in $\Lambda$CDM and that 
observed for the oldest objects in the Universe have recently 
been pointed out by \cite{Yu2014}.

And lest the reader come away with the sense that $\Lambda$CDM
and $R_{\rm h}=ct$ overlap so much that one should not worry
about their differences, we close this discussion by again pointing
out the most profound consequence of their disparity. As shown in
\cite{Melia2014c}, the horizon problem does not exist in $R_{\rm h}=ct$.
So whereas $\Lambda$CDM could not survive without inflation, the
real universe may have done without it, and the cosmological
data---particularly at high redshift---may be pointing in that
direction.

\acknowledgments
I am grateful to the anonymous referee for suggestions that have
led to improvements in the manuscript.
I am also grateful to Amherst College for its support through a John Woodruff Simpson
Lectureship, and to Purple Mountain Observatory in Nanjing, China, for its hospitality
while part of this work was being carried out. This work was partially supported by
grant 2012T1J0011 from The Chinese Academy of Sciences Visiting Professorships for
Senior International Scientists, and grant GDJ20120491013 from the Chinese State
Administration of Foreign Experts Affairs.


\begin{thebibliography}{}
\bibitem[\protect\citeauthoryear{Ade et al.}{2013}]{Ade2013} Ade, P.A.R. et al. (Planck Collaboration) 2013,
A\&A, in press (arXiv:1303.5062)
\bibitem[\protect\citeauthoryear{Bennett et al.}{2003}]{Bennett2003} Bennett, C. L. et al.\ 2003, ApJ, 583, 1
\bibitem[\protect\citeauthoryear{Bennett et al.}{2013}]{Bennett2013} Bennett, C. L. et al.\ 2013, ApJS, 208, 20
\bibitem[\protect\citeauthoryear{Copi et al.}{2009}]{Copi2009} Copi, C. J., Huterer, D., Schwarz, D. J. \& Starkman, G. D.\ 2009, 
MNRAS, 399, 295
\bibitem[\protect\citeauthoryear{Copi et al.}{2013}]{Copi2013} Copi, C. J., Huterer, D., Schwarz, D. J. \& Starkman, G. D.\ 2013, 
MNRAS in press (arXiv:1310.3831)
\bibitem[\protect\citeauthoryear{Guth}{1981}]{Guth1981} Guth, A. H.\ 1981, PRD, 23, 347
\bibitem[\protect\citeauthoryear{Guth et al.}{2013}]{Guth2013} Guth, A. H., Kaiser, D. I. \& Nomura, I.\ 2013, eprint arXiv:1312.7619
\bibitem[\protect\citeauthoryear{Ijjas et al.}{2013}]{Ijjas2013} Ijjas, A., Steinhardt, P. J. \& Loeb, A.\ 2013, PLB, 723, 261
\bibitem[\protect\citeauthoryear{Ijjas et al.}{2014}]{Ijjas2014} Ijjas, A., Steinhardt, P. J. \& Loeb, A.\ 2013, eprint arXiv:1402.6980
\bibitem[\protect\citeauthoryear{Kosowsky et al.}{2002}]{Kosowsky2002} Kosowsky, A., Milosavljevic, M. \& Jimenez, R.\ 2002,
PRD, 66,  063007
\bibitem[\protect\citeauthoryear{Linde}{1982}]{Linde1982} Linde, A.\ 1982, PLB, 108, 389
\bibitem[\protect\citeauthoryear{Maor et al.}{2001}]{Maor2001} Maor, I., Brustein, R. \& Steinhardt, P. J.\ 2001, PRL, 86, 6
\bibitem[\protect\citeauthoryear{Melchiorri et al.}{2003}]{Melchiorri2003} Melchiorri, A., Mersini, L., \"Odman, C. J. 
Trodden, M.\ 2003, PRD, 68, 043509
\bibitem[\protect\citeauthoryear{Melia}{2007}]{Melia2007} Melia, F.\ 2007, MNRAS, 382, 1917
\bibitem[\protect\citeauthoryear{Melia}{2012a}]{Melia2012a} Melia, F.\ 2012a, AJ, 144, id. 110
\bibitem[\protect\citeauthoryear{Melia}{2012b}]{Melia2012b} Melia, F.\ 2012b, JCAP, 09, 029
\bibitem[\protect\citeauthoryear{Melia}{2013a}]{Melia2013a} Melia, F.\ 2013, ApJ, 764, 72
\bibitem[\protect\citeauthoryear{Melia}{2013b}]{Melia2013b} Melia, F.\ 2013b, CQG, 30, 155007
\bibitem[\protect\citeauthoryear{Melia}{2014a}]{Melia2014a} Melia, F.\ 2014a, A\&A, 561, id A80
\bibitem[\protect\citeauthoryear{Melia}{2014b}]{Melia2014b} Melia, F.\ 2014b, AJ, 147, id 120
\bibitem[\protect\citeauthoryear{Melia}{2014c}]{Melia2014c} Melia, F.\ 2014c, A\&A, 553, id A76
\bibitem[\protect\citeauthoryear{Melia}{2014d}]{Melia2014d} Melia, F.\ 2014d, A\&A, 553, id A76
\bibitem[\protect\citeauthoryear{Melia \& Abdelqader}{2009}]{MeliaAbdelqader2009} Melia, F. \& Abdelqader, M.\ 2009, IJMP-D, 18, 1889
\bibitem[\protect\citeauthoryear{Melia \& Maier}{2013}]{MeliaMaier2013} Melia, F. \& Maier, R. S.\ 2013, MNRAS, 432, 2669
\bibitem[\protect\citeauthoryear{Melia \& Shevchuk}{2012}]{MeliaShevchuk2012} Melia, F. \& Shevchuk, A.S.H.\ 2012, MNRAS, 419, 2579
\bibitem[\protect\citeauthoryear{Misner \& Sharp}{1964}]{Misner1964} Misner, C. W. \& Sharp, D. H.\ 1964, Phys Rev, 136, 571
\bibitem[\protect\citeauthoryear{Perlmutter et al.}{1999}]{Perlmutter1999} Perlmutter, S. et al.\ 1999, ApJ, 517, 565
\bibitem[\protect\citeauthoryear{Riess et al.}{1998}]{Riess1998} Riess, A. G. et al.\ 1998, AJ, 116, 1009
\bibitem[\protect\citeauthoryear{Sachs \& Wolfe}{1967}]{Sachs1967} Sachs, R. K. \& Wolfe, A. M.\ 1967, ApJ, 147, 73
\bibitem[\protect\citeauthoryear{Spergel et al.}{2003}]{Spergel2003} Spergel, D. N. et al.\ 2003, ApJS, 148, 175
\bibitem[\protect\citeauthoryear{Suzuki et al.}{2012}]{Suzuki2012} Suzuki, N., Rubin, D., Lidman, C., et al.\ 2012, ApJ, 746, 85
\bibitem[\protect\citeauthoryear{Tegmark et al.}{2002}]{Tegmark2002} Tegmark, M., Hamilton, A.J.S. \& Xu, Y.\ 2002, MNRAS, 335, 887
\bibitem[\protect\citeauthoryear{Watson et al.}{2011}]{Watson2011} Watson, D. F., Berlind, A. A. \& Zentner, A. R.\ 2011,
ApJ, 738, id. 22
\bibitem[\protect\citeauthoryear{Wei et al.}{2013a}]{Wei2013a} Wei, J.-J., Wu, X.-F. \& Melia, F.\ 2013a, ApJ, 772, 43
\bibitem[\protect\citeauthoryear{Wei et al.}{2013b}]{Wei2013b} Wei, J.-J., Wu, X.-F., Melia, F. \& Maier, R. S.\ 2013b, AJ, in press 
\bibitem[\protect\citeauthoryear{Weinberg}{1972}]{Weinberg1972} Weinberg, S.\ 1972, Gravitation and Cosmology: 
Principles and Applications of the General Theory of Relativity (Wiley, New York)
\bibitem[\protect\citeauthoryear{Yu \& Wang}{2014}]{Yu2014} Yu, H. \& Wang, F. Y.\ 2014, EPJ-C, 74, id. 3090
\end{thebibliography}
\end{document}